\documentclass{article}
\usepackage[T1]{fontenc} 
\usepackage[utf8]{inputenc} 
\usepackage{url}
\def\authorname{F. Author, S. Author, and T. Author}

\def\papertitle{Helicality: An Isomap-based Measure of Octave Equivalence in Audio Data}

\usepackage{ismir-lbd}

\usepackage{lineno}
\usepackage[caption=false,font=footnotesize]{subfig}

\title{\papertitle}

\twoauthors
  {Sripathi Sridhar} {New York University\\New York, NY, USA}
  {Vincent Lostanlen} {Cornell Lab of Ornithology\\Ithaca, NY, USA}

\begin{document}

\maketitle
\begin{abstract}
Octave equivalence serves as domain-knowledge in MIR systems, including chromagram, spiral convolutional networks, and harmonic CQT. Prior work has applied the Isomap manifold learning algorithm to unlabeled audio data to embed frequency sub-bands in 3-D space where the Euclidean distances are inversely proportional to the strength of their Pearson correlations. However, discovering octave equivalence via Isomap requires visual inspection and is not scalable. To address this problem, we define "helicality" as the goodness of fit of the 3-D Isomap embedding to a Shepherd-Risset helix. Our method is unsupervised and uses a custom Frank-Wolfe algorithm to minimize a least-squares objective inside a convex hull. Numerical experiments indicate that isolated musical notes have a higher helicality than speech, followed by drum hits.
\end{abstract}

\section{Introduction}\label{sec:introduction}
Octave equivalence refers to the perceived consonance of any pair of tones whose frequency ratio is a power of two.
Although not universally shared across listeners, this phenomenon lies at the foundation of multiple theories of music around the globe.
For example, concepts such as \emph{Saptak} in Indian classical music or \emph{Tonnetz} in German classical music both rely on the hypothesis that notes one octave apart can be assigned to the same pitch class.

To explain computer-generated auditory paradoxes such as
the circularity of Shepard tones \cite{deutsch2010acoustics,risset1969jasa,pelofi2017philtrans}, music psychologists represent pitch on a 3-D helix (or, equivalently, a 2-D spiral) which makes a full turn at every octave, thereby aligning power-of-two harmonics  \cite{shepard1964jasa}.

In MIR, the earliest representation with octave equivalence is arguably the chromagram \cite{wakefield1999aspaai}, i.e., a constant-$Q$ transform (CQT) followed by summation across octaves.
More recently, the spiral scattering transform \cite{lostanlen2015dafx} applies a multivariable wavelet modulus operator before octave summation, thus improving discriminability while guaranteeing stability to octave transposition.

Another line of research considers octave equivalence as  domain-specific knowledge in deep learning.
Spiral convolutional networks \cite{lostanlen2016ismir} have receptive fields \emph{à trous} (``with holes'') spanning across octaves.
Comparable ideas are found in more recent publications, such as harmonic CQT \cite{bittner2017ismir}, folded CQT \cite{ducher2019ismir}, Fifthnet \cite{ohanlon2020icassp}, harmonic filters \cite{won2020icassp}, and harmonic convolutions \cite{zhang2020iclr}.

Despite the practical effectiveness of octave equivalence in various MIR tasks, the question of discovering octave equivalence directly from data has received less attention.
One exception is \cite{abdallah2003icabss}, which applied multidimensional scaling (MDS) to visualize mutual information between time--frequency atoms learned by independent component analysis (ICA).
More recently, \cite{lostanlen2020icassp} have applied the Isomap manifold learning algorithm \cite{tenenbaum2000science} to visualize Pearson correlations between CQT activations.


One shortcoming of methods \cite{abdallah2003icabss} and \cite{lostanlen2020icassp} is that they are purely illustrative: although they provide a scatter plot of time--frequency atoms in 3-D, the outcome is left to visual inspection, which hampers their scalability.
Our paper addresses this problem by proposing a geometrical criterion as to whether a 3-D scatter plot resembles a helix.
\let\thefootnote\relax\footnote{Companion website: \url{github.com/sripathisridhar/sridhar2020ismir}}

\section{Methods}\label{sec:method}

Given an unlabeled audio dataset of $N$ audio files, we use librosa v0.8.0 
to compute a CQT representation of every file with $Q$ = 24 bins per octave.
We restrict the dataset in the time domain to the loudest CQT frame in each audio file; and in the frequency domain, to the $J  = 3$ octaves of greatest variance.
This results in a matrix $\mathbf{X}$ with $P=24\times3=72$ rows and $N$ columns.

We extract squared Pearson correlations $\boldsymbol{\rho}^2[u,v]$ across all pairs of features, and apply the following formula:
\begin{equation}
    \mathbf{D}_{\boldsymbol{\rho}^2}[p,q]=
    \sqrt{-\dfrac{1}{2} \log \boldsymbol{\rho}^{2}[p,q] }
\end{equation}
to convert them into pseudo-Euclidean distances.
Following the methodology of Isomap, we use $\mathbf{D}_{\boldsymbol{\rho}^2}$ to compute a nearest-neighbor graph with $k=3$ neighbors per vertex.
With scikit-learn v0.20.0 
, we apply classical multidimensional scaling (MDS) to build an embedding space in which Euclidean distances approximate geodesic distances on the graph.
We refer to \cite{lostanlen2020icassp} for further details. 

Let $\mathbf{e}_m$ and $\lambda_m$ be the eigenvectors and eigenvalues resulting from MDS.
We rank eigenvalues in decreasing order without loss of generality.
We represent every subband $p$ by a point $\boldsymbol{y}[p] = (\mathbf{e}_1 [p], \mathbf{e}_2[p])$ on the plane. Let $\chi$ denote chroma. Then, we compute chroma centroids $\widetilde{\boldsymbol{y}}[\chi] = \dfrac{1}{J}\sum_{j=0}^{J-1}\boldsymbol{y}[\chi + Qj]$.
Our postulate is that if $\mathbf{X}$ has a property of octave equivalence, then the set of points $\mathcal{Y}=\{\widetilde{\boldsymbol{y}}[1] \ldots \widetilde{\boldsymbol{y}}[\chi]\}$ should form a circle.

We apply the Quickhull algorithm \cite{barber1996toms} to extract the convex hull of $\mathcal{Y}$, denoted by $\mathcal{H}$.
We denote by $\boldsymbol{c}_0$ the barycenter of vertices in $\mathcal{H}$.
Then, we fit a circle to $\mathcal{Y}$ by seeking a point $c$ inside $\mathcal{H}$ which minimizes the following objective:
\begin{equation}\label{eq:leastsquares}
    V_{\mathrm{circle}}(\boldsymbol{\boldsymbol{c}}) = \sum_{\chi=1}^{Q}\big\Vert \boldsymbol{c} - \widetilde{\boldsymbol{y}}[\chi] \big\Vert_2^2 -
    \dfrac{1}{Q}\left(\sum_{\chi=1}^{Q}\big\Vert \boldsymbol{c} - \widetilde{\boldsymbol{y}}[\chi]\big\Vert_2\right)^2,
\end{equation}
taken from \cite[Equation 4]{coope1993circle}.
In practice, we solve the problem above via a custom implementation of the Frank-Wolfe conditional gradient algorithm \cite{jaggi2013icml}, initialized at $\boldsymbol{c}_0$.

Likewise, we seek two parameters $a$ and $b$ such that the affine function $p\mapsto (a\times p+b)$ approximates the sequence $\boldsymbol{z} = \mathbf{e}_3$ by minimizing the following objective:
\begin{equation}
    V_{\mathrm{line}}(a, b) =
    \sum_{p=1}^{P}
    \big\Vert
    a \times p + b -
    \boldsymbol{z}[p]
    \big\Vert_2^2
\end{equation}
We solve the problem above by linear regression.

On the point cloud $\boldsymbol{\psi}[p] = (\boldsymbol{e}_1[p], \boldsymbol{e}_2[p], \boldsymbol{e}_3[p])$, we fit a helix based on the circle and line estimates, denoted by $\boldsymbol{\psi}^{'}[p]$ .
Then, we define helicality as the inverse of the mean squared Euclidean distance between the embedding points and the projected points in Equation \ref{eq:helicality}. This measures the deviation of the embedding from an ideal helix, denoting the extent of octave equivalence in the frequency content of the audio dataset.
\begin{equation} \label{eq:helicality}
    \boldsymbol{H} = \dfrac{1}{\dfrac{1}{P}\sum_{p=1}^{P}\big\Vert \boldsymbol{\psi}[p] - \boldsymbol{\psi}^{'}[p]\big\Vert_2^2},
\end{equation}

\section{Results}\label{sec:results}
We analyze three datasets: TinySOL (music) \cite{cella2020icmc}, which contains $2913$ recordings from $14$ instruments; ENST-drums (drums), from which we select the subset of $107$ isolated drum hits from $3$ drummers; and North Texas Vowel Dataset (speech), containing  $3190$ recordings of voiced vowels by $50$ American English speakers.

\begin{figure}
    \centering
    \subfloat[\centering{Music. Helicality: 0.54.}]{
    \includegraphics[width=0.45\linewidth]{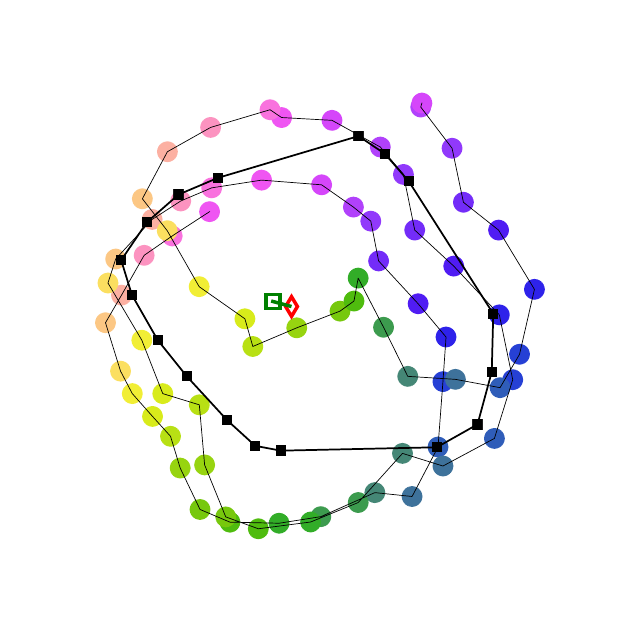}}
    \subfloat[\centering{Speech. Helicality: 0.30.}]{
    \includegraphics[width=0.45\linewidth]{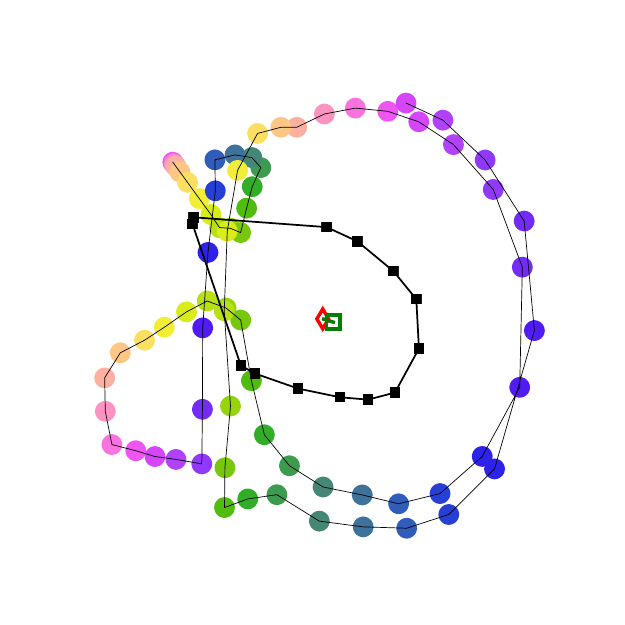}}
    \caption{Isomap embedding of music (left) and speech (right) data. The hue of the colored dots and the grey line denote pitch chroma and pitch height. The black squares and the solid black line represent the convex hull. The red diamond and green square respectively denote initial and final center estimate for circle fitting.}
    \label{fig:cross-dataset}
\end{figure}

On visual inspection, music data has a more helical embedding topology than speech data  in Fig. \ref{fig:cross-dataset}. Accordingly, speech data has a lower helicality score $H$ = 0.30, while music data has a higher score $H$ = 0.54. 

\begin{figure}
    \centering
    \subfloat[{Horn only. Helicality: 0.94.}]{
    \includegraphics[width=0.45\linewidth]{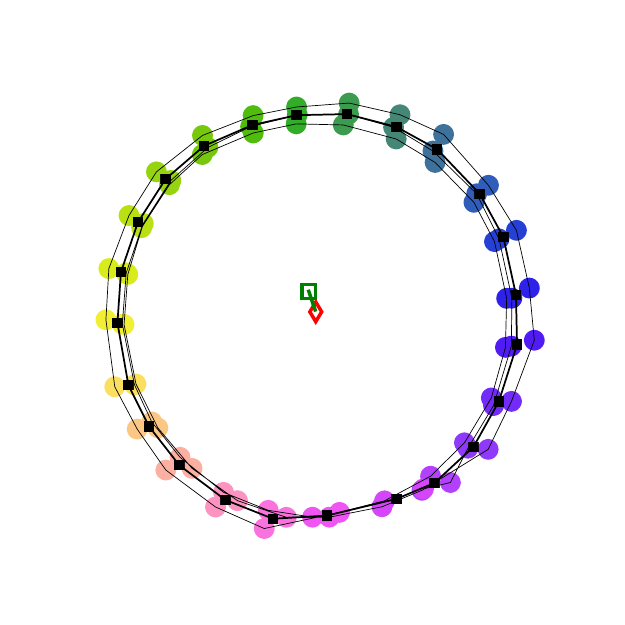}}
    \subfloat[\centering{Drums  only. Helicality: 0.28.}]{
    \includegraphics[width=0.45\linewidth]{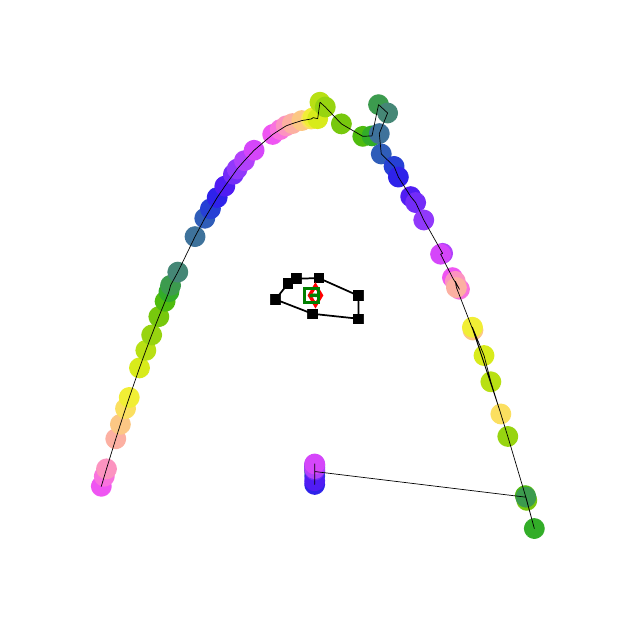}}
    \caption{Isomap embedding of isolated instruments: horn (left) and drums (right). See Figure 1 for legend.}
    \label{fig:cross-instrument}
\end{figure}

\begin{figure}
    \centering
    \includegraphics[width=\linewidth]{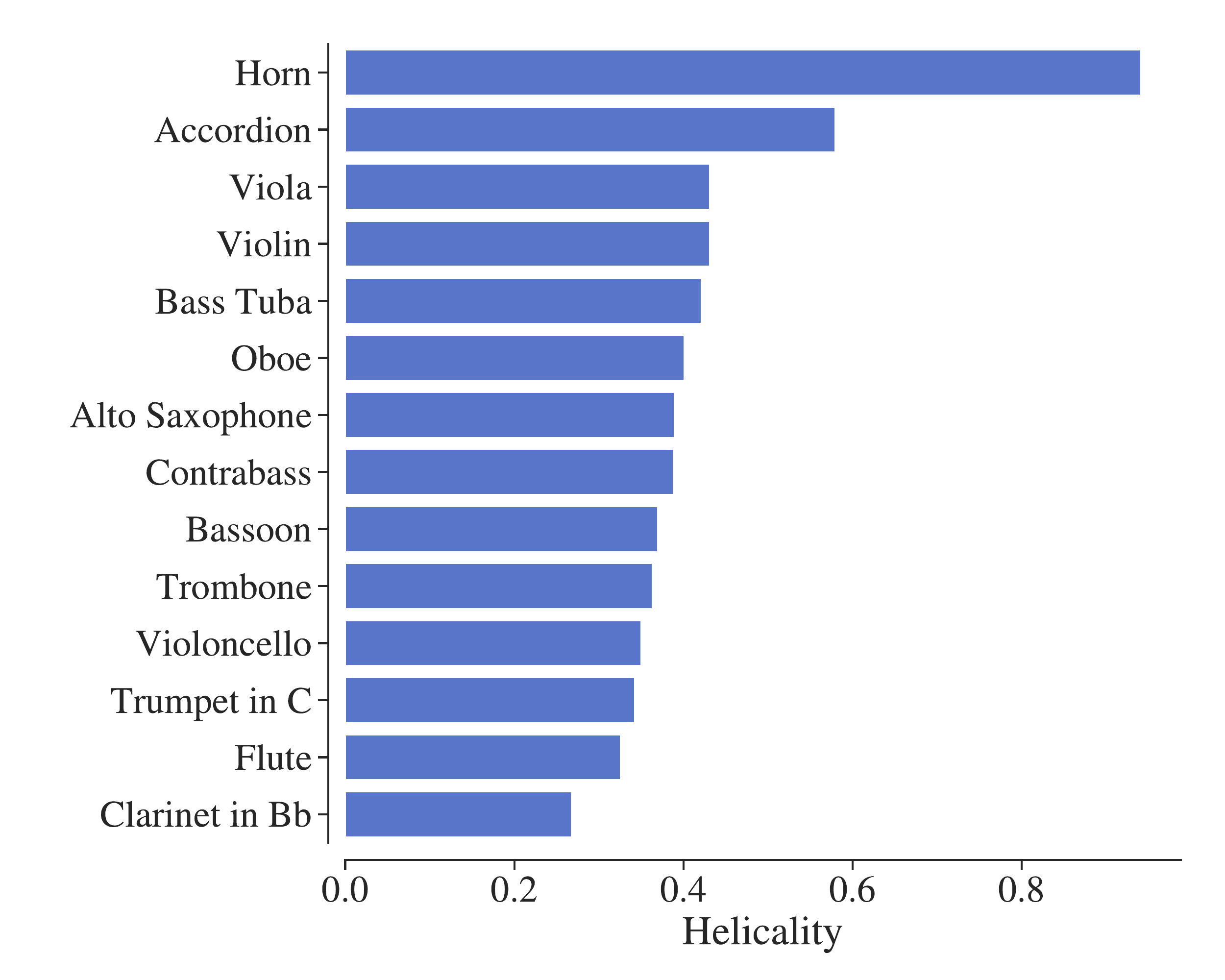}
    \caption{Helicality scores of TinySOL instrument classes.}
    \label{fig:tinysol_helicality}
\end{figure}

\emph{Horn} ($H$ = 0.94) produces the most helical embedding, seen in Fig. \ref{fig:cross-instrument}, followed by \emph{Accordion} ($H$ = 0.58).
Isolated drum hits have a lowest helicality score of $H$ = 0.28. 
Fig. \ref{fig:tinysol_helicality} compares different instrument classes.
Surprisingly, \emph{Trumpet} ($H$ = 0.34) has a low helicality score, despite its characteristic harmonic structure.

\section{Conclusion}
We have presented a method for fitting a helix to a 3-D manifold of CQT subbands.
Further research is needed to examine whether helicality matches human perception.

\section{Acknowledgment}
This work is supported by NSF award 1633259 (BirdVox).

\bibliography{sridhar2020ismir_lbd}

\begin{thebibliography}{10}
\providecommand{\url}[1]{#1}
\csname url@samestyle\endcsname
\providecommand{\newblock}{\relax}
\providecommand{\bibinfo}[2]{#2}
\providecommand{\BIBentrySTDinterwordspacing}{\spaceskip=0pt\relax}
\providecommand{\BIBentryALTinterwordstretchfactor}{4}
\providecommand{\BIBentryALTinterwordspacing}{\spaceskip=\fontdimen2\font plus
\BIBentryALTinterwordstretchfactor\fontdimen3\font minus
  \fontdimen4\font\relax}
\providecommand{\BIBforeignlanguage}[2]{{%
\expandafter\ifx\csname l@#1\endcsname\relax
\typeout{** WARNING: IEEEtran.bst: No hyphenation pattern has been}%
\typeout{** loaded for the language `#1'. Using the pattern for}%
\typeout{** the default language instead.}%
\else
\language=\csname l@#1\endcsname
\fi
#2}}
\providecommand{\BIBdecl}{\relax}
\BIBdecl

\bibitem{deutsch2010acoustics}
D.~Deutsch, ``The paradox of pitch circularity,'' \emph{Acoustics Today},
  vol.~7, pp. 8--15, 2010.

\bibitem{risset1969jasa}
J.-C. Risset, ``Pitch control and pitch paradoxes demonstrated with
  computer-synthesized sounds,'' \emph{Journal of the Acoustical Society of
  America}, vol.~46, no.~1A, pp. 88--88, 1969.

\bibitem{pelofi2017philtrans}
C.~Pelofi, V.~{de}~Gardelle, P.~Egr{\'e}, and D.~Pressnitzer, ``Interindividual
  variability in auditory scene analysis revealed by confidence judgements,''
  \emph{Philosophical Transactions of the Royal Society B: Biological
  Sciences}, vol. 372, no. 1714, p. 20160107, 2017.

\bibitem{shepard1964jasa}
R.~N. Shepard, ``Circularity in judgments of relative pitch,'' \emph{JASA},
  vol.~36, no.~12, pp. 2346--2353, 1964.

\bibitem{wakefield1999aspaai}
G.~Wakefield, ``{Mathematical Representation of Joint Time--chroma
  Distributions},'' in \emph{Proc. SPIE ASPAAI}, 1999.

\bibitem{lostanlen2015dafx}
V.~Lostanlen and S.~Mallat, ``Wavelet scattering on the pitch spiral,'' in
  \emph{Proc. DAFx}, 2015.

\bibitem{lostanlen2016ismir}
V.~Lostanlen and C.~E. Cella, ``Deep convolutional networks on the pitch spiral
  for musical instrument recognition,'' in \emph{Proc. ISMIR}, 2016.

\bibitem{bittner2017ismir}
R.~M. Bittner, B.~McFee, J.~Salamon, P.~Li, and J.~P. Bello, ``Deep salience
  representations for $f_0$ estimation in polyphonic music,'' in \emph{Proc.
  ISMIR}, 2017.

\bibitem{ducher2019ismir}
J.-F. Ducher and P.~Esling, ``{Folded CQT RCNN for Real-time Recognition of
  Instrument Playing Techniques},'' in \emph{Proc. ISMIR}, 2019.

\bibitem{ohanlon2020icassp}
K.~O’Hanlon and M.~B. Sandler, ``{The Fifthnet Chroma Extractor},'' in
  \emph{Proc. IEEE ICASSP}.\hskip 1em plus 0.5em minus 0.4em\relax IEEE, 2020.

\bibitem{won2020icassp}
M.~Won, S.~Chun, O.~Nieto, and X.~Serr\'{a}, ``Data-driven harmonic filters for
  audio representation learning,'' in \emph{Proc. IEEE ICASSP}.\hskip 1em plus
  0.5em minus 0.4em\relax IEEE, 2020.

\bibitem{zhang2020iclr}
Z.~Zhang, Y.~Wang, C.~Gan, J.~Wu, J.~B. Tenenbaum, A.~Torralba, and W.~T.
  Freeman, ``Deep audio priors emerge from harmonic convolutional networks,''
  in \emph{Proc. ICLR}, 2020.

\bibitem{abdallah2003icabss}
S.~A. Abdallah and M.~D. Plumbley, ``{Geometric ICA Using Nonlinear Correlation
  and MDS},'' in \emph{Proc. ICA/BSS}, 2003.

\bibitem{lostanlen2020icassp}
V.~Lostanlen, S.~Sridhar, B.~McFee, A.~Farnsworth, and J.~P. Bello, ``Learning
  the helix topology of musical pitch,'' in \emph{Proc. IEEE ICASSP}, 2020.

\bibitem{tenenbaum2000science}
J.~B. Tenenbaum, V.~De~Silva, and J.~C. Langford, ``A global geometric
  framework for nonlinear dimensionality reduction,'' \emph{Science}, vol. 290,
  no. 5500, pp. 2319--2323, 2000.

\bibitem{barber1996toms}
C.~B. Barber, D.~P. Dobkin, and H.~Huhdanpaa, ``The quickhull algorithm for
  convex hulls,'' \emph{TOMS}, vol.~22, no.~4, pp. 469--483, 1996.

\bibitem{coope1993circle}
I.~D. Coope, ``Circle fitting by linear and nonlinear least squares,''
  \emph{Journal of Optimization Theory and Applications}, vol.~76, no.~2, pp.
  381--388, 1993.

\bibitem{jaggi2013icml}
M.~Jaggi, ``{Revisiting Frank-Wolfe: Projection-free sparse convex
  optimization},'' in \emph{Proc. ICML}, 2013.

\bibitem{cella2020icmc}
C.~E. Cella, D.~Ghisi, V.~Lostanlen, F.~L{\'e}vy, J.~Fineberg, and Y.~Maresz,
  ``{OrchideaSOL: a dataset of extended instrumental techniques for
  computer-aided orchestration},'' in \emph{Proc. ICMC}, 2020.

\end{thebibliography}

\end{document}